\documentclass[aip, preprint]{revtex4-1}
\usepackage{bm, amsfonts, mathtools}
\usepackage{graphicx, color, wasysym}
\usepackage{textcomp}
\usepackage{amsmath, amssymb, amsthm, latexsym, epsfig}
\usepackage{appendix}

\usepackage{stackrel}

\usepackage{txfonts}

\DeclareMathOperator{\D}{d\!}
\DeclareMathOperator{\E}{e} 
\DeclareMathOperator{\I}{i}

\begin{document} 
\title[On the Sheffer-type polynomials related to the Mittag-Leffler functions {: applications to \ldots}]{On the Sheffer-type polynomials  related to the Mittag-Leffler functions{: applications to {fractional} evolution equations}}

\author{K. G\'{o}rska}
\email{katarzyna.gorska@ifj.edu.pl}
\altaffiliation[Also at ]{ENEA - Centro Ricerche Frascati, via E. Fermi, 45, IT-00044 Frascati (Roma), Italy} 
 
\author{A. Horzela}
 \email{andrzej.horzela@ifj.edu.pl}
 
\affiliation{H. Niewodniczanski Institute of Nuclear Physics, Polish Academy of Science (INP PAS), \\ ul. Radzikowskiego 152, PL-31342 Krak\'{o}w, Poland}

\author{K. A. Penson}
 \email{penson@lptl.jussieu.fr}
\affiliation{Laboratorie de Physique Theorique de la Mati\`{e}re Condens\'{e}e (LPTMC), CNRS UMR 7600, Sorbonne Universit\'{e}, Campus Pierre et Marie Curie, F-75005 Paris, France}

\author{G. Dattoli}
\email{giuseppe.dattoli@enea.it}
\affiliation{ENEA - Centro Ricerche Frascati, via E. Fermi, 45, IT-00044 Frascati (Roma), Italy}

\date{\today}

\begin{abstract}
We present two types of polynomials related to the Mittag-Leffler function namely the fractional Hermite polynomial and the Mittag-Leffler polynomial. The first modifies the Hermite polynomial and the second one is a refashioned Laguerre polynomial. The fractional Hermite and the Mittag-Leffler polynomials are used to solve {the Cauchy problems for} the fractional Fokker-Planck equation where the fractional derivative is taken in the Caputo sense with respect to time and/or space. The generating functions of these two kinds of polynomials are also calculated and they indicate that these polynomials belong to the Sheffer type.
\end{abstract}

\maketitle

\section{Introduction}\label{sec1}

There are several ways to define the Sheffer polynomials {\cite{IMSheffer39, SRoman84, SKhan15, PBlasiak06, GDattoli12}} among which the method by a generating function is the most common. The exponential generating function (EGF) of the Sheffer sequences $s_{n}(x)$, $n=0, 1, 2, \ldots$, can be expressed in terms of
\begin{equation}\label{8/10-1}
\sum_{n=0}^{\infty} \frac{\lambda^{n}}{n!} s_{n}(x) = A(\lambda) \E^{x B(\lambda)},
\end{equation}
where $A(0) = 1$ and $B(0) = 0$. If $B(\lambda) = \lambda$ we have the subclass of the Sheffer polynomials called the Appell polynomials \cite{SRoman84, SKhan15, Mansour}. To the family of the Sheffer polynomials belong, e.g., the Hermite and the Laguerre polynomials which are present in the solutions of the classical and quantum mechanical problems. For instance, it is well-known that the Hermite polynomials solve the heat equation\cite{DVWidder75} and anomalous diffusion \cite{RMetzler99}. They also appeared in the wave functions of the Schr\"{o}dinger equations with the harmonic oscillator potential \cite{LIShiff49} which in the three-dimensional case can be additionally written in terms of the associated Laguerre polynomials \cite{LIShiff49} connected with the hydrogen atom potential \cite{LIShiff49}. One of the important property of the Sheffer polynomials is that under the action of certain operators they behave like the monomial and by the monomiality principle create the ladder structure {\cite{PBlasiak06, KAPenson06, GDattoli12}}. From this point of view they are essential in theoretical physics. 

In this paper we will extend the class of the Sheffer polynomials by introducing two types of polynomials related to the Mittag-Leffler function (MLF) which are solutions of the fractional differential equation with fractional derivative over time and/or over space. First of these polynomials are called the fractional Hermite polynomials (fHP) and they are related to the (one-parameter, standard) MLF denoted by $E_{\alpha}(x)$, $\alpha > 0$, $x\in\mathbb{R}$. The next one, named the Mittag-Leffler polynomials (MLP) and denoted by $E_{\alpha, \,\beta}^{-n}(x)$, $n = 1, 2, \ldots$, $\alpha, \beta > 0$, and $x\in\mathbb{R}$, are the three-parameter MLF with the negative integer upper parameter. Both of these polynomials are related to the MLFs whose series definitions read
\begin{equation}\label{8/10-2}
E_{\alpha}(x) = \sum_{r=0}^{\infty} \frac{x^{r}}{\Gamma(1 + \alpha r)}, \qquad \qquad E_{\alpha, \,\beta}^{\,\gamma}(x) = \sum_{r=0}^{\infty} \frac{x^{r}}{\Gamma(\beta + \alpha r)} \frac{\Gamma(\gamma + r)}{\Gamma(\gamma)},
\end{equation}
and from which it can be easily seen that $E^{\gamma}_{\alpha, \,\beta}(x)$ goes to $E_{\alpha}(x)$ for $\gamma = \beta = 1$. In addition we observe that $E^{1}_{\alpha, \,\beta}(x) = E_{\alpha, \,\beta}(x) = \sum_{r=0}^{\infty} x^{r}/\Gamma(\alpha+\beta r)$, the Wiman function. It should be also added that the one-parameter MLF generalize the exponential function.

The one-parameter MLFs play a central role in the theory of fractional calculus which involves the integro-differential operator in the Caputo's form:  
\begin{equation}\label{8/10-3}
{^{C}\!D}_{t}^{\alpha} g(t) = \frac{1}{\Gamma(n-\alpha)} \int_{0}^{\,t} (t-y)^{n-\alpha-1} g^{(n)}(y) \D y, \qquad n-1 < \alpha < n,
\end{equation}
for $n=1, 2, \ldots$ and $g^{(n)}(y)$ being the $n$th derivative of $g(y)$ over $y$. The role of one-parameter MLF is similar to the one played by the exponential function in the conventional calculus with the derivative of integer order. Namely, the standard MLF is the eigenfunction of the fractional derivative in the Caputo sense: ${^{C}D}_{t}^{\alpha} E_{\alpha}(a t^{\alpha}) = a E_{\alpha}(a t^{\alpha})$, $a\in\mathbb{R}$, $\alpha>0$, and $t > 0$, where ${^{C}\!D}_{t}^{\alpha} g(t)$ is defined by Eq. \eqref{8/10-3}. {To prove that we use {Theorem 10.3} of Ref.\, \cite{HJHaubold09}, namely ${^{RL}D}_{t}^{\alpha} E_{\alpha}(a t^{\alpha}) = t^{-\alpha}/\Gamma(1-\alpha) + a E_{\alpha}(a t^{\alpha})$ in which ${^{RL}\!D}_{t}^{\alpha} g(t)$ is the Riemann-Liouville fractional derivative. Relation between the Caputo and the Riemann-Liouville fractional derivatives say that ${^{C}\!D}_{t}^{\alpha} g(t) = {^{RL}\!D}_{t}^{\alpha} g(t) - t^{-\alpha} g(0)/\Gamma(1-\alpha)$, see Ref. \cite[Eq. (2.254)]{IPodlubny99}.} 

The MLFs can be also studied as the non-Debye relaxation functions of the complex systems \cite{Garrappa, KGorska18a}. For the Cole-Cole relaxation function we have that $\psi_{\alpha, 1}(t) = E_{\alpha}[-(t/\tau)^{\alpha}]$ with $0 < \alpha < 1$ and $\tau$ is interpreted as the effective time constant which corresponds to relaxation time of whole sample. The three-parameters MLP appeared in the Havriliak-Negami relaxation function, i.e. $\psi_{\alpha, \, \beta}(t) = 1 - (t/\tau)^{\alpha\beta} E_{\alpha, 1 +\alpha\beta}^{\beta}[-(t/\tau)^{\alpha}]$ which for $\beta = 1$ goes to the relaxation function $\psi_{\alpha, 1}(t)$, i.e. the one-parameter MLF\, \cite{Garrappa, KGorska18a}.

The paper is organized as follows. In Sec. \ref{sec2} we introduce the two-variable fHPs which have the similar algebraical properties as two-variable Hermite polynomials. In Sec. \ref{sec2}  we also derive the exponential generating function and the forward shift operator appropriate for the fHPs. We have also shown that these polynomials solve the one-dimensional fractional Fokker-Planck equation only with diffusion term proportional to the second derivative over space. The MLP are discussed in Sec. \ref{sec3}. There, we  propose their operational form and we find their ordinary and exponential generating functions. We will show that these polynomials are used to solve the one-dimensional fractional Fokker-Planck equation in which the diffusion term (i.e. the second derivative over space) is replaced by the fractional Laguerre derivative ${^{\,C}\!D^{\,\beta}_{x} x\partial_{x}}$. These two types of polynomials belong to the Sheffer class so in Sec. \ref{sec3a} we will investigate the monomiality principle for them. The paper is concluded in Sec. \ref{sec4}.

{This work is an extension of considerations reported in Ref. \cite{ArXiv}.}

\ \\ \ \\ \ \\

\section{The fractional Hermite polynomials ${_{\alpha} H}_{n}(x, y)$}\label{sec2}

The fHP introduced in \cite[Eq. (8)]{KGorska12} {and \cite[Eq. (3)]{ArXiv}}reads
\begin{equation}\label{8/10-4}
{_{\alpha}H_{n}}(x, y) = n! \sum_{r=0}^{\lfloor n/2 \rfloor} \frac{x^{n-2 r}\; y^{r}}{(n-2r)! \Gamma(1+\alpha r)}, \qquad n=0, 1, 2, \ldots,
\end{equation}
and its first four terms, i.e. for $n=0, 1, 2$, and $3$, are equal to $1$, $x$, $x^{2} + 2y/\Gamma(1+\alpha)$, and $x^{3} + 6 xy/\Gamma(1+\alpha)$, respectively. According to the definition \eqref{8/10-4} the fHP differs from the Hermite polynomials in two variables $H_{n}(x, y) = n! \sum_{r=0}^{\lfloor n/2 \rfloor} x^{n-2r} y^{r}/[(n-2r)! r!]$ only by the gamma function in the denominator. Then it can be expected that the algebraic properties of ${_{\alpha}H_{n}}(x, y)$ and $H_{n}(x, y)$ should be similar. For example, as $H_{n}(x, y)$ can be presented in the form of standard Hermite polynomials $H_{n}(u)$ in the form $H_{n}(x, y) = (\I\! y^{1/2})^{n} H_{n}[x/(2\I\! y^{1/2})]$ then ${_{\alpha}H_{n}}(x, y)$ can be written using Eq. \eqref{8/10-4} as:
\begin{equation*}
{_{\alpha}H_{n}}(x, y) = (\I\! y^{1/2})^{n} {_{\alpha} H_{n}}[x/(2\I\! y^{1/2})].
\end{equation*}
The operational form of $H_{n}(x, y)$ reads $\exp(y\, \partial_{x}^{2}) x^{n}$\; \cite{Pino-dzielo} and it can be shown that
\begin{equation}\label{14/11-1}
{_{\alpha}H_{n}}(x, y) = E_{\alpha}(y \partial_{x}^{2}) x^{n}.
\end{equation} 
Eq. \eqref{14/11-1} can be immediately derived by applying the series form of the one-parameter MLF given through Eq. \eqref{8/10-2}. 

Taking into account the analogy in the defining sums between ${_{\alpha}H_{n}}(x, y)$ and $H_{n}(x, y)$ we can suppose that ${_{\alpha}H_{n}}(x, y)$ can be formally expressed via $H_{n}(x, y)$. It is done with the help of the umbral calculus which leads to the umbral image of the fHP:
\begin{equation}\label{8/10-5}
{_{\alpha} H_{n}}(x, y) = H_{n}(x, y\,d_{\rho}) M_{\alpha}(-\alpha\rho)\big\vert_{\rho=0},
\end{equation}
in which $M_{\alpha}(\sigma)= \Gamma(1-\sigma/\alpha)/\Gamma(1-\sigma)$ is the $\sigma$-th Stieltjes moment of the one-sided L\'{e}vy stable distribution $\varPhi_{\alpha}(u)$, $0 < \alpha < 1$\, \cite{HPollard46, KAPenson10, KGorska12a}. The symbol $d_{\rho}$ is called the umbral shift operator as it shifts the function $f(\rho)$ such that $d^{\,\sigma}_{\rho}: f(\rho) \mapsto f(\rho + \sigma)$. In the example studied here the shifted function $f(\rho)$ is equal to $M_{\alpha}(-\alpha\rho)$. Roughly speaking, Eq. \eqref{8/10-5} can be derived from Eq. \eqref{14/11-1} in which we use the umbral form of the one-parameter MLF involving the Stieltjes moments of $\varPhi_{\alpha}(u)$. According to Ref. {\cite{SilviaTh, KGorska18}} the one-parameter MLF can be rephrased as
\begin{equation}\label{8/10-6}
E_{\alpha}(a) = \E^{a d_{\rho}} M_{\alpha}(-\alpha\rho)\big\vert_{\rho = 0}, \qquad d_{\rho}^{\,\sigma} M_{\alpha}(-\alpha\rho)\big\vert_{\rho = 0} = M_{\alpha}(-\alpha\sigma).
\end{equation}
Thus, inserting Eq. \eqref{8/10-6} into Eq. \eqref{14/11-1} gives ${_{\alpha} H_{n}}(x, y) = \exp(y\, d_{\rho} \partial_{x}^{2})\, x^{n}\, M_{\alpha}(-\alpha\rho)\vert_{\rho = 0}$ and treating it as the operational form of $H_{n}(x, y)$ we obtain Eq. \eqref{8/10-5}.

From Eq. \eqref{8/10-5} the EGF of ${_{\alpha}H_{n}}(x, y)$ can be calculated which is equal to
\begin{equation}\label{19/06-9}
\sum_{n=0}^{\infty} \frac{\lambda^{n}}{n!}\, {_{\alpha}H_{n}}(x, y) = \E^{x \lambda} E_{\alpha}(y \lambda^{2}).
\end{equation}

\noindent
{\sc Proofs of Eq. \eqref{19/06-9}.} The immediate proof of Eq. \eqref{19/06-9} is obtained by applying the umbral representation of the fHP given by Eq. \eqref{8/10-5}. From that it turns out that 
\begin{equation*}
\text{RHS of Eq. \eqref{19/06-9}} = \sum_{n=0}^{\infty} \frac{\lambda^{n}}{n!}\, H_{n}(x, y\, d_{\rho}) M_{\alpha}(-\alpha\rho) \big\vert_{\rho = 0} = \E^{x \lambda} \E^{\,y  \lambda^{2} d_{\rho}} M_{\alpha}(-\alpha\rho), 
\end{equation*}
where we employed the EGF of $H_{n}(x, y)$ (see Ref. \cite{Pino-dzielo}). Using now Eq. \eqref{8/10-6} the umbral proof is concluded.  \\ 
Eq. \eqref{19/06-9} can be also proved by inserting the series form of the fHP into RHS of Eq. \eqref{19/06-9}. That gives
\begin{equation*}
\text{RHS of Eq. \eqref{19/06-9}} = \sum_{n=0}^{\infty} \sum_{r=0}^{\lfloor n/2 \rfloor} \frac{(\lambda x)^{n-2r}}{(n-2r)!}\, \frac{(y \lambda^{2})^{r}}{\Gamma(1 + \alpha r)} = \sum_{n=0}^{\infty} \frac{(\lambda x)^{n}}{n!} \sum_{r=0}^{\infty} \frac{(y \lambda^{2})^{r}}{\Gamma(1 + \alpha r)}
\end{equation*}
employing the summation over the triangle $0 \leq r \leq \lfloor n/2\rfloor$ in a different order. 

We can also derive the forward shift operators of the fHP on the first and the second arguments:
\begin{equation}\label{20/06-2}
\partial_{x}\, [{_{\alpha}H_{n}}(x, y)] = n\, [{_{\alpha}H_{n-1}}(x, y)] \quad \text{and} \quad {^{C\!}D^{\alpha}_{y}}\; [{_{\alpha}H_{n}}(x, y^{\alpha})] = n(n-1)\; [{_{\alpha}H_{n-2}}(x, y^{\alpha})]
\end{equation}
{\sc Umbral proof of Eqs. \eqref{20/06-2}.} The first equality in Eqs. \eqref{20/06-2} can be immediately shown by using Eq. \eqref{8/10-5} and the analogue relation of two variable Hermite polynomials due to which $\partial_{x} H_{n}(x, y) = n H_{n-1}(x, y)$. The second equality of Eq. \eqref{20/06-2} can be proved by using Eq. \eqref{8/10-4}, the facts that the MLF is the eigenfunction of fractional derivative in Caputo sense, and observing that $\partial^{\,2}_{x}$ commute with $E_{\alpha}(y \partial^{\,2}_{\,x})$. The calculations go as follows:
\begin{equation}\label{20/06-3}
{^{C\!}D^{\alpha}_{y}}\, [{_{\alpha}H_{n}}(x, y^{\alpha})] = [{^{C}D^{\alpha}_{y}}\, E_{\alpha}(y^{\alpha} \partial_{x}^{2})]\, x^{n} = E_{\alpha}(y^{\alpha} \partial_{x}^{2})\, \partial_{x}^{2} x^{n} = n(n-1) E_{\alpha}(y^{\alpha} \partial_{x}^{2})\, x^{n-2}.
\end{equation}
Employing once again Eq. \eqref{8/10-4} we obtain the RHS of the second equality of Eqs. \eqref{20/06-2}. 

The fHPs at $x=0$ are equal to
\begin{equation}\label{20/06-4}
{_{\alpha}H_{n}}(0, y) = \frac{n!\, y^{\,n/2}}{\Gamma(1 + \alpha n/2)} |\cos(n\pi/2)|,
\end{equation}
which is zero for odd $n$ and differs from zero for even $n$.\\
{\sc Umbral proof of Eq. \eqref{20/06-4}.} Eq. \eqref{8/10-5} enables one to present ${_{\alpha}H_{n}}(0, y)$ as the appropriate $H_{n}(0, y d_{\rho})$ acting on $M_{\alpha}(-\alpha\rho)$ at $\rho = 0$. Moreover, from zeros of the Hermite polynomials $H_{n}(0)$\, \cite{NIST} we can deduct that $H_{n}(0, b)$ is equal to
\begin{equation*}
H_{n}(0, b) = \frac{n!\, b^{\,n/2}}{\Gamma(1 + n/2)}  |\cos(n\pi/2)|.
\end{equation*}
From these two facts one obtains the form of ${_{\alpha}H_{n}}(0, y)$ given by Eq. \eqref{20/06-4}. \qed

\subsection{The solutions of fractional heat equations}

As is shown in Ref. \cite{KGorska12} the fHP are the formal solutions of the time-fractional Fokker-Planck equation:
\begin{equation}\label{8/10-7}
{^{C\!}D_{t}^{\alpha}} F_{\alpha}(x, t) = k_{\alpha} \partial_{x}^{2} F_{\alpha}(x, t), \qquad F_{\alpha}(x, 0) = f(x), \quad t > 0, \quad x\in\mathbb{R},
\end{equation}
with $\alpha\in(0, 1)$, $f(x) = x^{n}$, $k_{\alpha} > 0$, and the fractional derivative in the Caputo sense taken over time. In Ref. \cite{KGorska12} is also shown that if the initial condition $f(x)$ is the power series, that is $f(x) = \sum_{r=0}^{\infty} c_{r} x^{n}$, then the solution of Eq. \eqref{8/10-7} represents the series of the fHPs, namely $F_{\alpha}(x, t) = \sum_{r= 0}^{\infty} c_{r}\, {_{\alpha}H_{n}}(x, k_{\alpha} t^{\alpha})$. Here, we present two choices of $f(x)$ which enable us to sum up the series in $F_{\alpha}(x, t)$ such that the final results contain the fHP. For this purpose we solve Eq. \eqref{8/10-7} for the two initial conditions: {\bf (i)} $f_{\rm\bf(i)}(a; x) = H_{n}(x, a)$ and {\bf (ii)} $f_{\rm{\bf{(ii)}}}(a; x) = {_{\alpha}H_{n}}(x, a)$.  In both these cases the parameter $a$ is real. Moreover, we will use the notation ${^{j\!}F_{k; \alpha}}(x, t)$ in which the superscript $j = \rm{\bf{(i)}}, \rm{\bf{(ii)}}$, is related to the below itemized points {\bf(i)} and {\bf (ii)}, whereas the subscript $k=1, 2$ informs us about solution calculated in two different ways.

\noindent
{\bf (i)} For $f_{\rm\bf{(i)}}(a; x) = H_{n}(x, a)$ in which we apply the sum form of $H_{n}(x, a)$ we get 
\begin{align}\label{20/06-6}
\begin{split}
{^{\rm{\bf{(i)\!}}}F}_{1; \alpha}(x, t) & = E_{\alpha}(t^{\,\alpha}k_{\alpha} \partial_{x}^{\,2}) H_{n}(x, a) = n! \sum_{r=0}^{\lfloor n/2 \rfloor} \frac{a^{r}}{r! (n - 2 r)!} E_{\alpha}(t^{\,\alpha}k_{\alpha} \partial_{x}^{\,2}) x^{n - 2r} \\
& = n! \sum_{r=0}^{\lfloor n/2 \rfloor} \frac{a^{r} {_{\alpha}H_{n - 2r}}(x, t^{\alpha} k_{\alpha})}{r!\, (n - 2 r)!}.
\end{split}
\end{align}
On the other hand the action of $E_{\alpha}(t^{\alpha} k_{\alpha} \partial_{x}^{\,2})$ on the initial condition $H_{n}(x, a)$ can be also calculated in an alternative way, namely by applying the operational form of $H_{n}(x, a)$ given above in Eq. \eqref{14/11-1}. That leads to ${^{\rm{\bf{(i)\!}}}F}_{2; \alpha}(x, t) = E_{\alpha}(t^{\alpha} k_{\alpha} \partial_{x}^{2}) \exp(a \partial_{x}^{2}) x^{n}$. Then from the umbral image of the MLF we have ${^{\rm{\bf{(i)\!}}}F}_{2; \alpha}(x, t) = \exp[(a + t^{\alpha} k_{\alpha} d_{\rho})\, \partial_{x}^{\,2}] x^{n} M_{\alpha}(-\alpha\rho)\vert_{\rho = 0}$. The action of $\exp(-b \partial_{x}^{2})$ with $b = a + t^{\alpha} k_{\alpha} d_{\rho}$ on the smooth function $x^{n}$ can be computed with the help of the Gauss-Weierstrass transform (see Ref. \cite{QR}). This leads to:
\begin{align}\label{21/06-5}
\begin{split}
{^{\rm{\bf{(i)\!}}}F}_{2; \alpha}(x, t) & = \frac{1}{2 [\pi(a + t^{\alpha} k_{\alpha} d_{\rho})]^{1/2}} \int_{-\infty}^{\infty} \exp\left[-\frac{(x-\xi)^{2}}{4(a + t^{\,\alpha} k_{\alpha} d_{\rho})}\right] \xi^{\,n} \D\xi\; M_{\alpha}(-\alpha\rho)\big\vert_{\rho=0} \\
& = H_{n}(x, a + t^{\alpha} k_{\alpha} d_{\rho})\; M_{\alpha}(-\alpha\rho)\big\vert_{\rho=0},
\end{split}
\end{align}
where Eq. (2.3.15.9) of Ref. \cite{APPrudnikov-v1}, i.e. $\int_{-\infty}^{\infty} x^{n} \exp(-p x^{2} -  q x) \D x = \sqrt{\pi/p} \,(-2p)^{-n} \exp[q^{2}/(4p)]\, {H}_{n}(q, p)$, is employed.

Eqs. \eqref{20/06-6} and \eqref{21/06-5} are calculated in two different ways. Nevertheless they are solutions of the {\em same} fractional Fokker-Planck equation with the {\em same} initial condition. That suggests that they are equal to each other:
\begin{equation}\label{21/06-6}
{H_{n}}(x, a + t^{\alpha} k_{\alpha} d_{\rho})\; M_{\alpha}(-\alpha\rho)\big\vert_{\rho=0} = n! \sum_{r=0}^{\lfloor n/2 \rfloor} \frac{a^{r} {_{\alpha}H_{n - 2r}}(x, t^{\alpha} k_{\alpha})}{r!\, (n - 2 r)!}.
\end{equation}
{\sc Proof of Eq. \eqref{21/06-6}.} To prove Eq. \eqref{21/06-6} we use the finite sum form of two-variable Hermite polynomials in which we present the second argument $(a + t^{\alpha} k_{\alpha} d_{\rho})^{r}$ as the Newton binomial sum, namely $r! \sum_{k=0}^{r} a^{k} (t^{\alpha} k_{\alpha} d_{\rho})^{r-k}/[k! (r-k)!]$. Hence, we find
\begin{align}\label{26/06-5}
\begin{split}
H_{n}(x, a + t^{\alpha} k_{\alpha} d_{\rho}) M_{\alpha}(-\alpha \rho)\vert_{\rho = 0} & = n! \sum_{r=0}^{\lfloor n/2 \rfloor} \frac{x^{n -2r}}{(n - 2r)! r!} \sum_{k=0}^{r} \binom{r}{k} a^{k} (t^{\alpha} k_{\alpha})^{r-k} d^{r-k}_{\rho} M_{\alpha}(-\alpha\rho)\big\vert_{\rho=0} \\
& = n! \sum_{r=0}^{\lfloor n/2 \rfloor} \frac{x^{n -2r}}{(n - 2r)! r!} \sum_{k=0}^{r} \binom{r}{k} a^{k} (t^{\alpha} k_{\alpha})^{r-k} \frac{(r-k)!}{\Gamma[1+\alpha(r-k)]},
\end{split}
\end{align}
where the action of umbral operator $d^{r-k}_{\rho}$ on $M_{\alpha}(-\alpha\rho)|_{\rho=0}$ gives $(r-k)!/\Gamma[1 + \alpha(r-k)]$. The double sums over $r=0, 1, \ldots, \lfloor n/2\rfloor$, and over $k = 0, 1, \ldots, r$, reflect the commutativity property over the triangle. They can be changed into other finite double sum where one is over $k = 0, 1, \ldots, \lfloor n/2 \rfloor$, and other is over $r = k, k+1, \ldots, \lfloor n/2 \rfloor$. Then, Eq. \eqref{26/06-5} reads
\begin{equation*}
\text{RHS of Eq. \eqref{26/06-5}} = n! \sum_{k=0}^{\lfloor n/2 \rfloor} \frac{a^{k}}{k!} \sum_{r=k}^{\lfloor n/2 \rfloor} \frac{x^{n-2r} (t^{\alpha} k_{\alpha})^{r-k}}{(n-2r)!} \Gamma[1 + \alpha(r-k)].
\end{equation*}
Setting $r-k = j$, where $j = 0, 1, \ldots, \lfloor (n-2k)/2\rfloor$ we get
\begin{equation}\label{26/06-7}
\text{RHS of Eq. \eqref{26/06-5}} = n! \sum_{k=0}^{\lfloor n/2 \rfloor} \frac{a^{k}}{k!} \sum_{j=0}^{\lfloor (n-2k)/2 \rfloor}  \frac{x^{(n-2k) - 2j} (t^{\alpha} k_{\alpha})^{j}}{[(n-2k) - 2j]! \Gamma(1+\alpha j)}.
\end{equation}
In Eq. \eqref{26/06-5}, recognizing ${_{\alpha} H_{n}}(x, t^{\alpha} k_{\alpha})$ which is multiplied by $(n - 2k)!$ in the sum over $j$ concludes the proof. 

\noindent
{\bf (ii)} The formal solution of Eq.~\eqref{8/10-7} for $f_{\rm{\bf{(ii)}}}(a; x) = {_{\alpha}H_{n}}(x, a)$ is given through ${^{\rm{\bf{(ii)\!}}}F_{\alpha}}(x, t) = E_{\alpha}(t^{\alpha}k_{\alpha}\partial_{x}^{\,2}) {_{\alpha}H_{n}}(x, a)$. Making the similar calculation like in previous point {\rm{\bf{(i)}}} but now employing the series form of ${_{\alpha}H_{n}}(x, a)$ we have
\begin{align*}
\begin{split}
{^{\rm{\bf{(ii)\!}}}F}_{1; \alpha}(x, t) & = n! \sum_{r=0}^{\lfloor n/2\rfloor} \frac{a^{r}}{(n-2r)! \Gamma(1+\alpha r)} E_{\alpha}(t^{\alpha} k_{\alpha} \partial_{x}^{2}) x^{n-2r} \\
& = n! \sum_{r=0}^{\lfloor n/2\rfloor} \frac{{_{\alpha}H_{n - 2r}}(x, t^{\,\alpha}k_{\alpha})\, a^{r}}{(n - 2r)! \Gamma(1 + \alpha r)}.
\end{split}
\end{align*}
A different way to compute ${^{\rm{\bf{(ii)\!}}}F}_{2; \alpha}(x, t)$ is to apply the operational representation of the initial condition. That implies ${^{\rm{\bf{(ii)\!}}}F}_{2; \alpha}(x, t) = E_{\alpha}(t^{\,\alpha} k_{\alpha} \partial_{x}^{\,2}) E_{\alpha}(a \partial_{x}^{\,2}) x^{n}$. As shown in Refs. {\cite{SilviaTh, KGorska18, GDattoli17}} the product of two MLF, i.e. $E_{\alpha}(x) E_{\alpha}(y)$, is equal to $E_{\alpha}(x \oplus_{\alpha} y)$, where the function $(x \oplus_{\alpha} y)^{n}$ means
\begin{equation}\label{21/06-2}
(x \oplus_{\alpha} y)^{n} = \sum_{r=0}^{n} \binom{n}{r}_{\alpha} x^{n-r} y^{r} \qquad \text{with} \quad \binom{n}{r}_{\alpha} = \frac{\Gamma(1+\alpha n)}{\Gamma(1+ \alpha r) \Gamma[1 + \alpha(n-r)]}.
\end{equation}
Using Eq. \eqref{21/06-2} it can be shown that $(a x \oplus_{\alpha} a y)^{n} = a^{n} (x \oplus_{\alpha} y)^{n}$. Thus, ${^{\rm{\bf{(ii)\!}}}F}_{2; \alpha}(x, t)$ becomes
\begin{align}\label{21/06-3}
\begin{split}
{^{\rm{\bf{(ii)\!}}}F}_{2; \alpha}(x, t) & = E_{\alpha}(t^{\alpha} k_{\alpha} \partial_{x}^{\,2} \oplus_{\alpha} a \partial_{x}^{\,2}) x^{n} = E_{\alpha}[(t^{\alpha} k_{\alpha} \oplus_{\alpha} a) \partial_{x}^{\,2}] x^{n} \\
& = {_{\alpha}H_{n}}(x, t^{\alpha} k_{\alpha} \oplus_{\alpha} a).
\end{split}
\end{align}

The solutions ${^{\rm{\bf{(ii)\!}}}F}_{1; \alpha}(x, t)$ and ${^{\rm{\bf{(ii)\!}}}F}_{2; \alpha}(x, t)$ are calculated by applying two various procedures so they should be equal to each other. It means that we can write
\begin{equation}\label{21/06-4}
{_{\alpha} H_{n}}(x, t^{\alpha} k_{\alpha} \oplus_{\alpha} a) = n! \sum_{r=0}^{\lfloor n/2 \rfloor} \frac{{_{\alpha}H_{n-2r}}(x, t^{\,\alpha} k_{\alpha})\, a^{r}}{(n-2r)! \Gamma(1+\alpha r)}.
\end{equation}
{\sc Proof of Eq. \eqref{21/06-4}.} Taking the finite sum  representation of the fHP given by Eq. \eqref{8/10-4} and employing the definition of $(x \oplus_{\alpha} y)^{r}$ given by Eq. \eqref{21/06-2} we get
\begin{equation*}
{_{\alpha} H_{n}}(x, t^{\alpha} k_{\alpha} \oplus_{\alpha} a) = n! \sum_{r=0}^{\lfloor n/2\rfloor} \frac{x^{n-2r}}{(n-2r)! \Gamma(1+\alpha r)} \sum_{k=0}^{r} \binom{r}{k}_{\alpha} (t^{\alpha} k_{\alpha})^{r-k} a^{k}.
\end{equation*}
Now, proceeding analogously like in the proof of Eq. \eqref{21/06-6}, namely using the commutativity property of finite double sums over the triangle and, thereafter, changing the summation index we will conclude the proof. \qed

\section{The Mittag-Leffler polynomials}\label{sec3}

The MLPs in two variables {\cite{ArXiv}} are defined as follows
\begin{equation}\label{24/06-1}
E^{-n}_{\alpha, \,\beta}(x, y) = \sum_{r=0}^{n} \binom{n}{r} \frac{(-x)^{r} y^{n-r}}{\Gamma( \beta + \alpha r)}, 
\end{equation}
where $\alpha, \beta > 0$. {(Remark that Eq. \eqref{24/06-1} comes from Eq. \eqref{8/10-2} in which $\gamma = - n$ and the use of $\Gamma(-\mu)\Gamma(1+\mu) = -\pi/\sin(\pi \mu)$).} For $n=0$ it is equal to $1/\Gamma(\beta)$ which is obviously equal to one when $\beta = 1, 2$ and it differs from one for other values of $\beta > 0$. Thus, in the literature Refs. \cite{JDEKonhauser67, HMSrivastava82} one considers the regularized version of MLP, called the generalized Konhauser polynomials $\tilde{Z}_{n}^{\beta}(\alpha; x, y) = \Gamma(\beta + \alpha n)\, E^{-n}_{\alpha, \,\beta}(x^{\alpha}, y)/n!$\, \cite{KGorska}. For $y=1$ and positive integer $\alpha$ it goes into the standard Konhauser polynomials \cite{JDEKonhauser67, HMSrivastava82}. For $\alpha = 1$ and $\beta > 0$ the MLPs can be reduced to the associated Laguerre polynomials in two variables \cite{DBabusci17}. Like it is for the fHPs, the MLPs in two variables can be written as the MLP in one real variable as follows
\begin{equation*}
E^{-n}_{\alpha, \,\beta}(x, y) = y^{n} E^{-n}_{\alpha, \,\beta} (x/y, 1) = y^{n}E^{-n}_{\alpha, \,\beta}(x/y).
\end{equation*}
This formula obtains immediately from Eq. \eqref{24/06-1}. 

The operational version of the MLP for $\beta = 1$ is given as
\begin{equation}\label{24/06-4}
E^{\,-n}_{\alpha, 1}(x^{\,\alpha}, y) = \E^{-\,(y/\alpha)\, {^{C\!}D^{\,\alpha}_{x}}\, x\, \partial_{x}} \frac{(-1)^{n} x^{\,\alpha n}}{\Gamma(1 + \alpha n)},
\end{equation}
where $(-1/\alpha) {^{C\!}D^{\,\alpha}_{x}} x \partial_{x}$ is called the fractional Laguerre derivative and $ {^{C\!}D^{\alpha}_{x}}$ denotes the fractional derivative in the Caputo sense defined by Eq. \eqref{8/10-3}. The fractional Laguerre derivative for $\alpha\to 1$ tends to the ordinary Laguerre derivative $-\partial_{x} x \partial_{x}$ used in Refs. \cite{DBabusci17, KAPenson09}. \\
{\sc Proof of Eq. \eqref{24/06-4}.} The RHS of Eq. \eqref{24/06-4} means that
\begin{equation}\label{24/06-5}
E^{\,-n}_{\alpha, 1}(x^{\,\alpha}, y) = \sum_{r=0}^{\infty} \frac{(-y/\alpha)^{r}}{r!}\, \Big[{^{C\!} D_{x}^{\alpha}}\, x\, \partial_{x}\Big]^{r} \frac{(-1)^{n} x^{\,\alpha n}}{\Gamma(1 + \alpha n)}.
\end{equation}
Using ${^{C\!} D_{x}^{\alpha}} x^{\gamma} = \Gamma(1+\gamma)\, x^{\gamma-\alpha}/\Gamma(1 + \gamma - \alpha)$ for $0 < \alpha < 1$ we derived 
\begin{equation}\label{24/06-6}
\Big[{^{C\!} D_{x}^{\,\alpha}} x\, \partial_{x}\Big]^{r} x^{\gamma} = \gamma(\gamma - \alpha)\ldots [\gamma - (r-1)\alpha]\, x^{\gamma - r\alpha} =  \frac{\Gamma(1+\gamma)}{\Gamma(1+\gamma - r \alpha)} x^{\gamma - r\alpha}
\end{equation}
for fixed values of $r$. This equality can be proved by using the mathematical induction method. Substituting Eq. \eqref{24/06-6} in which $\gamma = \alpha n$ into the RHS of Eq. \eqref{24/06-5} we get Eq. \eqref{24/06-1}, i.e. the MLP. \qed 

Eq. \eqref{24/06-4} enables us to calculate their ordinary generating function (OGF) and EGF as
\begin{align}\label{25/06-2}
\sum_{n=0}^{\infty} \lambda^{n} E^{-n}_{\alpha, 1}(x^{\,\alpha}, y) & = \E^{-(y/\alpha) {^{C\!}D^{\,\alpha}_{x}} x \partial_{x}} E_{\alpha}(- \lambda x^{\alpha}) \\ \sum_{n=0}^{\infty} \frac{\lambda^{n}}{n!} E^{-n}_{\alpha, 1}(x^{\alpha}, y) & = \E^{-(y/\alpha) {^{C\!}D^{\,\alpha}_{x}} x \partial_{x}}  W_{\alpha, 1}(-\lambda x^{\alpha}) 
\label{25/06-6}
\end{align}
where $W_{\alpha, \,\mu}(x) = \sum_{r=0}^{\infty} x^{r}/[r! \Gamma(\mu + \alpha r)]$ is known as the Wright function \cite{KST}. Those equations confirm the result in Eqs. (2.1) and (2.2) of \,\cite{DBabusci17} written down here in Eqs. \eqref{9/10-2} in {\em Remark}. \\
{\sc Proof of Eqs. \eqref{25/06-2} and \eqref{25/06-6}.} Substituting the operational form of MLP into the OGF of $E^{-n}_{\alpha, 1}(x^{\alpha}, y)$ we straightforwardly get Eq. \eqref{25/06-2}. The proof of Eq. \eqref{25/06-6} goes analogically like the presented proof of Eq. \eqref{25/06-2} so it is omitted here.  \qed

\noindent
{\em Remark.} The generating functions \eqref{25/06-2} and \eqref{25/06-6} are equal to Eqs. (2.1) and (2.2) of Ref. \cite{DBabusci17}. That is
\begin{equation}\label{9/10-2}
\sum_{n=0}^{\infty} \lambda^{n} E_{\alpha, \,\beta}^{-n}(x, y) = \frac{1}{1-\lambda y} E_{\alpha, \,\beta}\left(-\frac{\lambda x}{1-\lambda y}\right) \quad \text{and} \quad \sum_{n=0}^{\infty} \frac{\lambda^{n}}{n!} E_{\alpha, \,\beta}^{-n}(x, y) = \E^{\lambda y} W_{\alpha, \,\beta}(-\lambda x),
\end{equation}
where their derivations can be found.\\
{\sc Proof of Eqs. \eqref{9/10-2}.}  The proof of Eqs. \eqref{9/10-2} involving the umbral representation of the of the MLF presented by Ref. \cite{DBabusci17}, in which $E^{-n}_{\alpha, \,\beta}(x, y) = c_{z}^{\beta-1}(y - xc_{z}^{\alpha})^{n} [\Gamma(1+z)]^{-1}\vert_{z=0}$ where $c_{z}^{\mu} [\Gamma(1+z)]^{-1}\vert_{z=0} = [\Gamma(1+\mu)]^{-1}$. Here, we prove these equations in the conventional way, that is by inserting the series form of the MLPs in the RHS of Eqs. \eqref{9/10-2}. Proceeding in this way the RHS of OGF reads
\begin{align}\label{25/06-3}
\begin{split}
\sum_{n=0}^{\infty} \lambda^{n} E^{-n}_{\alpha, \,\beta}(x, y) & = \sum_{n=0}^{\infty} \sum_{r=0}^{n}  \binom{n}{r} \frac{(-\lambda x)^{r} (\lambda y)^{n-r}}{\Gamma(\beta+\alpha r)} \\
& = \sum_{r=0}^{\infty} \sum_{n=r}^{\infty} \binom{n}{r} \frac{(-\lambda x)^{r} (\lambda y)^{n-r}}{\Gamma(\beta+\alpha r)}.
\end{split}
\end{align}
Now, setting $k = n-r$ for $k = 0, 1, 2, \ldots$ we get that Eq. \eqref{25/06-3} implies
\begin{equation*}
\sum_{n=0}^{\infty} \lambda^{n} E^{-n}_{\alpha, \,\beta}(x, y) = \sum_{r=0}^{\infty} \frac{(-\lambda x)^{r}}{r! \Gamma(\beta+\alpha r)} \sum_{k=0}^{\infty} \frac{(k+r)!}{k!} (\lambda y)^{k}.
\end{equation*}
The sum over $k$ is the definition of the hypergeometric function ${_{1}F_{0}}(1 + r; \lambda y)$ multiplied by $r!$ and it can be simplified as $r!(1-\lambda y)^{-1-r}$. Cancelling $r!$ and using the series form of $E_{\alpha}(\cdot)$ we derive the first equation in Eqs. \eqref{9/10-2}. \\
The EGF can be achieved in the similar way in which the calculation goes as follows:
\begin{align*}
\begin{split}
\sum_{n=0}^{\infty} \frac{\lambda^{n}}{n!} E^{\,-n}_{\alpha, \,\beta}(x, y) & = \sum_{n=0}^{\infty} \sum_{r=0}^{n} \frac{(\lambda y)^{n-r}}{(n-r)!} \frac{(-\lambda x)^{r}}{r! \Gamma(\beta + \alpha r)} \\
& = \sum_{r=0}^{\infty} \frac{(-\lambda x)^{r}}{r! \Gamma(\beta + \alpha r)} \sum_{n=r}^{\infty} \frac{(\lambda y)^{n-r}}{(n-r)!}.
\end{split}
\end{align*}
Making the same set of the summation index $k=n-r$ as in Eq. \eqref{25/06-3} and applying the series definition of the Wright function given below Eq. \eqref{25/06-6} we obtain the EGF given by Eq. \eqref{9/10-2}. \qed

\subsection{Solutions of the fractional differential equation}

The formal solution of the differential equation with the space-fractional Laguerre derivative reads
\begin{equation}\label{9/10-3}
\partial_{t} G_{\alpha}(x, t) = - (b/\alpha)\, {^{C\!}D_{x}^{\alpha} x \partial_{x}}\, G_{\alpha}(x, t), \qquad G_{\alpha}(x, 0) = g(x), \quad t > 0, \quad x\in\mathbb{R}
\end{equation}
and it can be written as the action of the evolution operator $\exp[-(t b/\alpha){^{C\!}D_{x}^{\alpha} x \partial_{x}}]$ on the initial condition $g(\alpha; x)$. If $g_{n}(\alpha; x) = (-x^{\alpha})^{n}/\Gamma(1 + \alpha n)$ then we have the operational form of MLP given by Eq. \eqref{24/06-4}. For the series form of the initial condition, i.e. $g(\alpha; x) = \sum_{r=0}^{\infty} c_{r} g_{r}(\alpha; x)$ the solution $G_{\alpha}(x, t)$ can be expressed as the series of the MLPs, i.e. $G_{\alpha}(x, t) = \sum_{r=0}^{\infty}c_{r} E_{\alpha,\,\beta}^{-r}(x, t)$. Their special cases are given by the generating functions \eqref{9/10-2}. Namely, it is the OGF of $E_{\alpha, 1}^{-n}(x^{\alpha}, t)$ for $c_{r} = 1$ and their EGF  for $c_{r} = 1/r!$.

If we add the time-fractional derivative in Caputo sense to Eq. \eqref{9/10-3} then we will have the fractional differential equation in the form 
\begin{equation}\label{9/10-4}
{^{C\!}D_{t}^{\,\beta}} G_{\alpha, \,\beta}(x, t) = - (b/\alpha)\, {^{C\!}D_{x}^{\alpha} x \partial_{x}}\, G_{\alpha, \,\beta}(x, t), \qquad G_{\alpha, \,\beta}(x, 0) = \gamma(x).
\end{equation}
whose formal solution, according to \cite[Eq. (6)]{KGorska12}, is given by $E_{\beta}[-(t^{\,\beta} b/\alpha) {^{C\!}D_{x}^{\alpha} x \partial_{x}}]\gamma(x)$.  Using the integral representation of the one-parameter MLF \, \cite{HPollard48, KGorska12, EBarkai01, GDattoli17}, due to which 
\begin{equation}\label{9/10-5}
E_{\beta}(- u t^{\,\beta}) = \int_{0}^{\infty} n_{\beta}(s, t) \E^{-s u} \D u, \qquad n_{\beta}(s, t) = \frac{1}{\beta} \frac{t}{s^{1+1/\beta}} \varPhi_{\beta}(t s^{-1/\beta}),
\end{equation}
where $\varPhi_{\beta}(u)$ denotes the one-sided L\'{e}vy stable distribution, the formal solution of Eq. \eqref{9/10-4} can be written as 
\begin{equation}\label{14/11-2}
G_{\alpha, \,\beta}(x, t) = \int_{0}^{\infty} n_{\beta}(s, t) \E^{- s\, (t^{\,\beta} b/\alpha)\, {^{C\!}D_{x}^{\alpha} x \partial_{x}}} \gamma(x) \D s.
\end{equation}
Now, we specify Eq. \eqref{14/11-2} for the given initial conditions $\gamma(x)$. \\
\noindent
\rm{\bf{(iii)}} Eq. \eqref{14/11-2} for ${\gamma_{\rm{\bf{(iii)}}}}(x) = g_{n}(\alpha; x)$ has the form
\begin{equation}\label{26/06-11}
{^{\rm{\bf (iii)}}G}_{\alpha, \,\beta}(n; x, t) = \sum_{r=0}^{n} \frac{n!}{r!} \frac{(-x^{\alpha})^{r} (b t^{\,\beta})^{n-r}}{\Gamma(1+\alpha r) \Gamma[1 + \beta(n-r)]}.
\end{equation}
{\sc Proof of Eq. \eqref{26/06-11}.} From Eq. \eqref{24/06-4} it follows that Eq. \eqref{14/11-2} for $\gamma_{\rm{\bf{(iii)}}}(x) = g_{n}(\alpha; x)$ reads
\begin{equation}\label{26/06-11a}
{^{\rm{\bf (iii)}}G}_{\alpha, \,\beta}(n; x, t) = \int_{0}^{\infty} n_{\beta}(s, t) E^{-n}_{\alpha, 1}(x^{\,\alpha}, b s) \D s, 
\end{equation}
which after employing the series representation of $E^{-n}_{\alpha, 1}(x^{\,\alpha}, b s)$ has the form
\begin{equation}\label{26/06-12}
{^{\rm{\bf (iii)}}G}_{\alpha, \,\beta}(n; x, t) = \sum_{r=0}^{n} \binom{n}{r} \frac{(-x^{\alpha})^{r} b^{n-r}}{\Gamma(1 + \alpha r)} \int_{0}^{\infty} n_{\beta}(s, t)\, s^{n-r} \D s.
\end{equation}
To compute the integral in Eq. \eqref{26/06-12}, i.e. the Stieltjes moment problem of $n_{\beta}(s, t)$, we need to involve the explicit form of $n_{\beta}(s, t)$ given by Eq. \eqref{9/10-5}. Moreover, setting $t s^{-1/\beta} = u$ we represent these Stieltjes moments as the well-known Stieltjes moments of the one-sided L\'{e}vy stable distribution $\varPhi_{\beta}(u)$ \cite{KAPenson10}, i.e. 
\begin{equation}\label{26/06-13}
\int_{0}^{\infty} n_{\beta}(s, t)\, s^{n-r} \D s = t^{\,\beta(n-r)} \int_{0}^{\infty} u^{-\beta (n-r)} \varPhi_{\beta}(u) \D u, 
\end{equation}
which, according to Eq. \eqref{8/10-5}, is equal to $M_{\beta}[-\beta(n-r)]$ and
\begin{equation}\label{26/06-13}
\int_{0}^{\infty} n_{\beta}(s, t)\, s^{n-r} \D s = \frac{(n-r)!\, t^{\,\beta(n-r)}}{\Gamma[1+\beta(n-r)]}.
\end{equation}
Inserting it into Eq. \eqref{26/06-12} we constructed the RHS of Eq. \eqref{26/06-11}. \qed 

\noindent
{\bf (iv)} Taking the Wright function as the initial condition, this is ${\gamma_{\rm{\bf{(iv)}}}}(y; x) = W_{\alpha, 1}(-y x^{\alpha})$, we have 
\begin{equation}\label{26/06-14}
{^{\rm{\bf{(iv)}}}G}_{\alpha, \,\beta}(y; x, t) = W_{\alpha, 1}(-y x^{\,\alpha}) E_{\beta}(b y t^{\,\beta}). 
\end{equation}
{\sc Proof of Eq. \eqref{26/06-14}.} The formal solution given through Eq. \eqref{14/11-2} with ${\gamma_{\rm{\bf{(iv)}}}}(y; x) = W_{\alpha, 1}(-y x^{\alpha})$ gives 
\begin{align}\label{10/10-1}
\begin{split}
{^{\rm{\bf{(iv)}}}G}_{\alpha, \,\beta}(x, t) &= \int_{0}^{\infty} n_{\beta}(s, t) \E^{-(sb/\alpha)\, {^{C\!}D_{x}^{\alpha}}x\partial_{x}} W_{\alpha, 1}(-y x^{\alpha}) \D s\\
& = \left[\int_{0}^{\infty} n_{\beta}(s, t) \E^{s b y} \D s\right]\, W_{\alpha, 1}(-y x^{\alpha}) .
\end{split} 
\end{align}
In Eq. \eqref{10/10-1} we used the OGF given by Eq. \eqref{25/06-6}. That implies that the Wright function $W_{\alpha, 1}(-y x^{\alpha})$ is independent from the integration variable $s$. Then applying the series form of the exponential function $\exp(s b y)$ we obtain the second equality in Eq. \eqref{10/10-1} which contains the moment problem of $n_{\beta}(s, t)$ calculated in Eq.  \eqref{26/06-13}. That concludes the proof. \qed

{In conclusion we should quote alternative studies of other evolution-type fractional equations, carried out by E. Barkai and collaborators: fractional Langevin \cite{BB}, and fractional Feynman-Kac \cite{CB, TCB, WDB} equations.}

\section{Sheffer polynomials}\label{sec3a}

The fHP and MLP are the examples of the Appell polynomials i.e. the Sheffer polynomials of Eq. \eqref{8/10-1} for $B(\lambda) = \lambda$ \cite{Mansour}. In the first case the EGF of Eq. \eqref{19/06-9} enables one to claim that ${_{\alpha}H_{n}}(x, y)$ is the Appell polynomial in $x$ variable whereas $y$ is treated as the parameter. Similar situation is with $E_{\alpha, \,\beta}^{-n}(x, y)$; the difference is that the MLP is the Appell polynomial in $y$ variable and $x$ is treated as the parameter. Generally, the Sheffer sequence $s_{n}(x)$ can be calculated via the action of the exponentiation the operator $q(x) \D/\D x + v(x)$ on the arbitrary smooth function $f(x)$:
\begin{equation*}
\exp\left\{\lambda\left[q(x) \frac{\D}{\D x} + v(x)\right]\right\} f(x) = h(\lambda, x) f(T(\lambda, x)),
\end{equation*}
in which the auxiliary functions, this is $q(x)$, $v(x)$, $h(\lambda, x)$, and $T(\lambda, x)$, are defined with the help of the functions $A(\lambda)$ and $B(\lambda)$ that appeared in the EGF of Eq. \eqref{8/10-1}:
\begin{align}\label{15/10-2}
\begin{split}
q(x) = B'[B^{-1}(x-1)] \,, \qquad & \quad v(x) = \frac{A'[B^{-1}(x-1)]}{A[B^{-1}(x-1)]}\,, \\
T(\lambda, x) = B[\lambda + B^{-1}(x-1)] + 1\,, \qquad & \quad h(\lambda, x) = \frac{A[\lambda + B^{-1}(x-1)]}{A[B^{-1}(x-1)]}\,.
\end{split}
\end{align}
For Eq. \eqref{15/10-2} we refer to Eqs.(42)-(45) of Ref. \cite{KAPenson06}. The prime symbol above denotes the first derivative with respect to the argument and $u^{-1}(x)$ is the compositional inverse of $u(x)$. For the Appell polynomials $B(\lambda) = B^{-1}(\lambda) = \lambda$ hence $q(x) = 1$ and $T(\lambda, x) = \lambda + x$. Then, for the Appell polynomial only $v(x)$ and $h(\lambda, x)$ should be found. For the polynomials worked out in this paper the quantities $v(x)$ and $h(\lambda, x)$ can be obtained explicitly. \\
\noindent
{\bf (v)} From Eq. \eqref{19/06-9} we see that in the case of the fHP we have $A(\lambda; y) = E_{\alpha}(y \lambda^{2})$. At the beginning we calculate derivative of $A(\lambda; y)$ with respect to $\lambda$:
\begin{equation}\label{15/10-3}
A'(\lambda; y) = 2y\lambda \frac{\D}{\D\tau} E_{\alpha}(\tau)\Big\vert_{\tau=y \lambda^{2}} = \frac{2}{\alpha \lambda} E_{\alpha, 0}(y \lambda^{2}),
\end{equation}
where we use \cite[Eq. (5.2)]{HJHaubold09}. The auxiliary functions $v(x)$ and $h(\lambda, x)$ computed from Eq. \eqref{15/10-2} are equal to 
\begin{equation}\label{15/10-4}
v_{\rm{\bf (v)}}(x; y) = \frac{2}{\alpha(x-1)} \frac{E_{\alpha, 0}[y(x-1)^{2}]}{E_{\alpha}[y(x-1)^{2}]} \qquad \text{and} \qquad h_{\rm{\bf (v)}}(\lambda, x; y) = \frac{E_{\alpha}[y(\lambda + x-1)^{2}]}{E_{\alpha}[y(x-1)^{2}]}.
\end{equation}

\noindent
{\bf (vi)} From the EGF of MLP given by Eq. \eqref{9/10-2} we have that $A(\lambda; x) = W_{\alpha, \,\beta}(-\lambda x)$ whose derivative obtained from the series form of the Wright function is given as $A'(\lambda; x) = -x W_{\alpha, \,\beta + \alpha}(-\lambda x)$. Thus, the auxiliary functions $v(y; x)$ and $h(\lambda, y; x)$ are equal to
\begin{equation}\label{16/10-2}
v_{\rm{\bf (vi)}}(y; x) = - x\frac{W_{\alpha, \,\beta+\alpha}[-x(y-1)]}{W_{\alpha, \,\beta}[-x(y-1)]} \qquad \text{and} \qquad h_{\rm{\bf (vi)}}(\lambda, y; x) = \frac{W_{\alpha, \,\beta}[-x(\lambda + y -1)]}{W_{\alpha, \,\beta}[-x(y-1)]}.
\end{equation}

{Note that in {\bf (v)} $x$ is a variable and $y$ is a parameter, whereas in {\bf (vi)}  their roles are reversed: $y$ is a variable and $x$ is a parameter .}

\medskip
\noindent
{\em Monomiality principle {\cite{GDattoli07, GDattoli00, GDattoli00a}}.} 
\noindent
As mentioned in Sec. \ref{sec1} the monomiality principle mimics the ladder structure and thus it plays the crucial role especially in Quantum Mechanics. In Ref.\, \cite[Theorem 2.5.3]{SRoman84} it is exhibited that the EGF of the Appell sequence $w_{n}(x)$ can be also written as
\begin{equation*}
\sum_{r=0}^{\infty} \frac{\lambda^{n}}{n!} w_{n}(u) = \frac{1}{g(\lambda)}\exp(u \lambda)
\end{equation*}
which is satisfied for all $x\in \mathbb{C}$. The function $g(\lambda)$ can be conceived as the (formal) power series, namely $g(\lambda) = \sum_{r=0}^{\infty} g_{r}\, \lambda^{r}/r!$, such that $g_{0} \neq 0$. The monomiality principle \cite{GDattoli} fulfilled by the Appell sequences $w_{n}(x)$ is built from the ladder operators $P$ and $M$:
\begin{equation*}
P\, w_{n}(u) = n\, w_{n-1}(u) \qquad \text{and} \qquad M\, w_{n}(u) = w_{n+1}(u),
\end{equation*}
for which $[P, M] = PM - MP = 1$, see \,\cite[Theorem 2.5.6]{SRoman84} or \,\cite[Eqs. (1.16)-(1.18)]{SKhan15}. The lowering $P$ and raising $M$ operators are uniquely determined by the function $g(\lambda)$ as follows \cite{GDattoli07}:
\begin{equation}\label{11/10-3}
P = D \qquad \text{and} \qquad M = X - \frac{g'(D)}{g(D)},
\end{equation}
where $X$ and $D$ are the analogues of position and momentum operators in Quantum Mechanics, see \,\cite{SKhan15, PBlasiak06}. For our polynomial sets these quantities can be exactly calculated.\\
For the fHP $g(\lambda; y)$ is given though $1/A(\lambda; y)$ and, here, it is equal to $g(\lambda; y) = 1/E_{\alpha}(y\lambda^{2})$. Its derivative calculated with the help of Eq. \eqref{15/10-3} leads to $g'(\lambda; y) = -2/(\alpha \lambda) E_{\alpha, 0}(y\lambda^{2})/[E_{\alpha}(y\lambda^{2})]^{2}$. Thus, Eq. \eqref{11/10-3} yields $M = X + v_{(\rm{\bf v})}(D+1; y)$ where $v_{(\rm{\bf v})}(\cdot, -)$ is given by Eq. \eqref{15/10-4}.
%
For the MLP $g(\lambda; x) = 1/W_{\alpha, \,\beta}(-\lambda x)$ whose derivative is computed by using the derivative of Wright function and is equal to $g'(\lambda; x) = x W_{\alpha, \, \beta + \alpha}(-\lambda x)/[W_{\alpha,\,\beta}(-\lambda x)]^{2}$. The raising operator $M$ also reads $M = X + v_{(\rm{\bf vi})}(D+1; x)$ with $v_{(\rm{\bf vi})}(\cdot; -)$ defined by Eq. \eqref{16/10-2}.

\section{Conclusion}\label{sec4}

In this paper we introduced two kinds of polynomials which are connected with the MLF, widely applied in the fractional dynamics and the relaxation processes. First kind of polynomials has a definition similar but not the identical to the Hermite polynomials of two variables and, hence, they are called by the fHP. The second one comes from the three-parameters MLF with the negative upper parameter. The latter polynomials generalize the known in mathematical literatures Konhauser polynomials which in turn are related to the Laguerre polynomials. We show that both families of these polynomials can be used to solve the fractional Fokker-Planck equation with the fractional derivative with respect to time and/or space, taken in the Caputo sense. A few examples of Cauchy problems involving these polynomials are also studied in the paper. The exponential generating functions of the fractional Hermite and Mittag-Leffler polynomials indicate that they belong to the family of Appell polynomials and are members of the Sheffer family. Thus, with the known rules \cite{KAPenson06} we found for them the monomiality principle and, then, the appropriate ladder operators.

\begin{acknowledgments}
K.G. and A.H. were supported by the Polish National Center for Science (NCN) research grant OPUS12 no.~UMO-2016/23/B/ST3/01714.~K. G. acknowledges support from the Polish National Agency for Academic Exchange (NAWA) Programme Bekker, project no. PPN/BEK/2018/1/00184. 

K.G. would like to thank ENEA - the Research Center in Frascati (Roma) for warm hospitality.
\end{acknowledgments}

\end{document}